\documentclass[12pt,english]{article}
\def\be{\begin{equation}}
\def\eq{\end{equation}}
\usepackage{graphicx}
\usepackage{amsmath}
\usepackage{amssymb}
\begin{document}
\title{ A non-perturbative proof of Bertrand's theorem}
\author{F C Santos
\footnote{e-mail: filadelf@if.ufrj.br}\\ V Soares\footnote{e-mail: vsoares@if.ufrj.br} \\ A C Tort
\footnote{e-mail: tort@if.ufrj.br.}\\
Instituto de F\'{\i}sica
\\
Universidade Federal do Rio de Janeiro\\
Caixa Postal 68.528; CEP 21941-972 Rio de Janeiro, Brazil}
\maketitle

\begin{abstract}
We discuss an alternative non-perturbative proof of Bertrand's theorem  that leads in a concise way directly to the two allowed fields: the newtonian and the isotropic harmonic oscillator central fields. 
\end{abstract}
\section{Introduction}

In 1873, J. Bertrand\cite{bertrand} published a short but important paper in which he proved that there are of only two central fields for which all orbits radially bounded are closed, namely: The newtonian field and the isotropic harmonic oscillator field. Because of this additional degenerescency it is no wonder that the properties of those two fields have been under close scrutiny since Newton's times. Newton addresses to the isotropic harmonic oscillator in proposition X Book I, and to the inverse-square law in proposition XI \cite{Newton}. Newton shows that both fields give rise to an elliptical orbit with the difference that in the first case the force is directed towards the geometrical centre of the ellipse and in the second case the force is directed to one of the foci. Bertrand's result, also known as Bertrand's theorem, continues to fascinate old and new generations of physicists interested in classical mechanics and unsurprisingly papers devoted to it continue to be produced and published. Bertrand's proof concise and elegant and contrary to what one may be led to think by a number of perturbative demonstrations that can be found in modern literature, textbooks and papers on the subject, it is fully non-perturbative. As examples of perturbative demonstrations the reader can consult references \cite{tikochinsky, Brown, Zarmi}. We can also find in the literature demonstrations that resemble the spirit of Bertrand's original work as for example \cite{Arnold}. As far as the present authors are aware of all those demonstrations have a restrictive feature, i.e., they set a limit on the number of possibilities of the existence of central fields with the property mentioned above to a finite number and finally show explicitly that  among the surviving possibilities only two, the newtonian and the isotropic harmonic oscillator, are  really possible.  
%Recently its 100th anniversary was celebrated in a meeting @.

In his paper, Bertrand proves initially by taking into consideration the equal radii limit that a central force $f(r)$ acting on a point-like body able of generating radially bounded orbits must necessarily be of the form

\[ f\left( r\right) =\kappa\,r^{\left(1/m^2-3\right)} ,\] 
where $r$ is the radial distance to center of force, $\kappa$ is a constant and  $m$ a rational number. Next, making use of this particular form of the law of force and considering also an additional limiting condition, Bertrand finally shows that only for  $m=1$ and $m=1/2$, which correspond to Newton's gravitational law of force

\[f\left( r\right) =-\frac{\kappa}{r^{2}} ,\]
and to the isotropic harmonic oscillator law of force

\[f\left( r\right) = - \kappa\,r,\] 
respectively, we can have orbits with the properties stated in the theorem. 
However, we can also prove that for these laws of force all bounded orbits are closed. 

Here we offer an alternative non-perturbative proof of Bertrand's theorem  that leads in a more concise way directly to the two allowed fields.
\section{Bertrand's theorem}
In a central field one can introduce a potential function $V\left(
r\right) $, through the property  

\be
\mathbf{f}=-\mathbf{\nabla}\, V\left( r\right) .
\eq
in such a way that the mechanical energy of a point-like body of mass $\mu$

\be E=\frac{\mu}{2} v^{2} +V\left(
r\right) ,
\eq 
is  conserved. For radially bounded orbits there are two extreme radii  $r_{\max }$ e $r_{\min }$, the so called apsidal points $r_{a}$, that are determined by the condition $\dot{r}_{a}=0$, and between which the particle oscillates indefinitely. Moreover, the conservation of the angular momentum of the particle under the action of a central field obliges the motion to take place on a fixed plane and allows the introduction of the effective potential
\begin{equation}
U(r)=V\left( r\right) +\frac{\ell^{2}}{2 \mu^{2}},  \label{potencial_eft}
\end{equation}
with the help of which it is possible to reduce this problem to an equivalent unidimensional one. This procedure can be found in several textbooks at the undergraduate and graduated level, see for example \cite{goldstein}. In terms of the effective potential orbits radially bounded are characterised by apsidal distances  $r_{\max }$ e $r_{\min }$ that satisfy the condition $E=U\left( r_{a}\right)$. Evidently there is an intermediate  point $r_{0}$ where the effective potential has a minimum that satisfies
\begin{equation}
U^{\prime }\left( r_{0}\right) =V^{\prime }\left( r_{0}\right) -\frac{\ell^{2}}{%
\mu r_{0}^{3}}=0 . \label{minimodeU}
\end{equation}
The angular displacement of the particle between two successive apsidal points, 
the apsidal angle $\Delta \theta _{a}$, is determined by
\begin{equation}
\Delta \theta _{a}=\int_{r_{\min }}^{r_{\max }}\displaystyle{\frac{\ell}{\mu r^{2}}}\frac{dr}{%
\sqrt{\displaystyle{\frac{2}{\mu}}\left[ E-U\left( r\right) \right] }}.  \label{delta_ang}
\end{equation}
%%%%%%%%%%%%%%%%%%%%%%%%%%%%%%%%%%%%%%%%%%%%%%%%
\begin{figure}[!h]
\begin{center}
\includegraphics{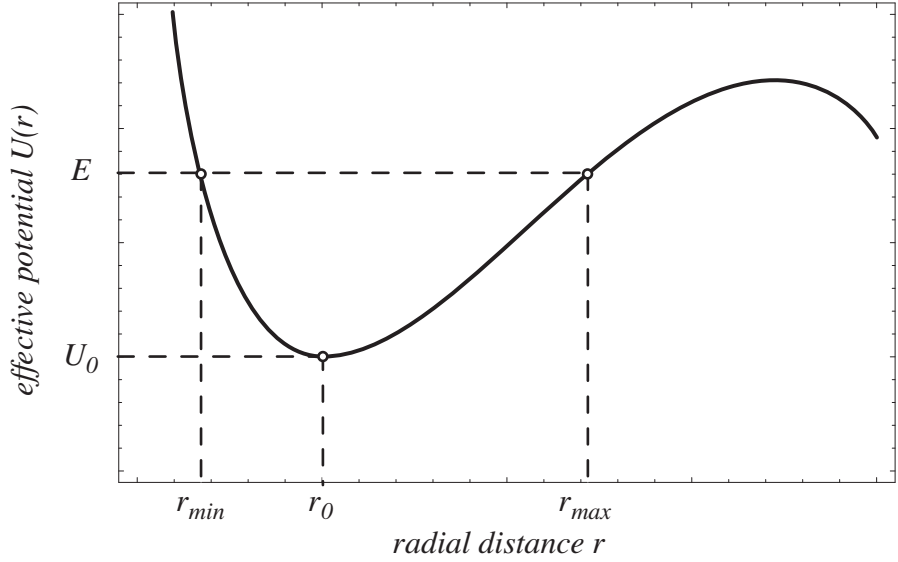}
\caption{General form of the effective potential energy.} 
\label{Potential1}
\end{center}
\end{figure}
%%%%%%%%%%%%%%%%%%%%%%%%%%%%%%%%%%%%%%%%%%%%%%%%

By considering the effective potential $ U $ as the independent variable
and by making use of the inverse function $
r\left( U\right) $, Tikochinsky\cite{tikochinsky} produced a very ingenious proof of Bertrand's theorem.  The inversion of the equation (\ref
{potencial_eft}), however, is not possible in all the domain on which the radial coordinate $r$ is defined because the function is not one-to-one in the field of the real numbers. To circumvent this difficulty we define two one-to-one branches of the function $U\left( r\right) $, namely, one to the left and the other to the right of the point $r_{0}$. Then we introduce the inverse functions $r_{1}=r_{1}\left( U\right) $ and $r_{2}=r_{2}\left( U\right) $, defined to the left and to the right of the point $r_{0},$ respectively, see Figure \ref{Potential1} .

We express initially the angular displacement, equation when the particle moves from the point of minimum radial distance $r_{\min }$ to the point $r_{0}$ in terms of the variable  $U$
\begin{equation}
\Delta \theta _{1}=\int_{E}^{U_{0}}\frac{\ell}{\mu r_{1}^{2}}\frac{dr_{1}}{dU}%
\frac{dU}{\sqrt{\displaystyle{\frac{2}{\mu}}\left[ E-U\right] }}=-\int_{E}^{U_{0}}\frac{\ell}{\mu}%
\frac{d}{dU}\left( \frac{\ell}{r_{1}}\right) \frac{dU}{\sqrt{\displaystyle{\frac{2}{\mu}}\left[
E-U\right] }}  . \label{delta_ang1}
\end{equation}
By the same token we will also have 
\begin{equation}
\Delta \theta _{2}=\int_{U_{0}}^{E}\frac{\ell}{\mu r_{2}^{2}}\frac{dr_{2}}{dU}%
\frac{dU}{\sqrt{\displaystyle{\frac{2}{\mu}}\left[ E-U\right] }}=-\int_{U_{0}}^{E}-\frac{\ell}{\mu}%
\frac{d}{dU}\left( \frac{1}{r_{2}}\right) \frac{dU}{\sqrt{\displaystyle{\frac{2}{\mu}}\left[
E-U\right] }}  , \label{delta_ang2}
\end{equation}
for the angular displacement from $r_{0}$ to the point of maximum radial distance $r_{2}$. Upon adding up equations (\ref{delta_ang1}) and (\ref
{delta_ang2}) we obtain the angular displacement between two successive apsidal points
\begin{equation}
\Delta \theta _{a}=\int_{U_{0}}^{E}F\left( U\right) \frac{dU}{\sqrt{E-U}},
\label{delta_angT}
\end{equation}
where
\begin{equation}
F\left( U\right) =\frac{\ell}{\sqrt{2\mu}}\frac{d}{dU}\left( \frac{1}{r_{1}}-%
\frac{1}{r_{2}}\right) .  \label{FdeU}
\end{equation}

Equation (\ref{delta_angT}) is Abel's integral equation the solution of which can be found, for example, in Landau's well known book on classical mechanics \cite{landau}. A beautiful and straightforward solution of this equation is the one by Oldham and Spanier \cite{oldham}. Abel's solution reads
\begin{equation}
\frac{1}{r_{1}}-\frac{1}{r_{2}}=\frac{\sqrt{2\mu}}{\pi \ell}\int_{U_{0}}^{U}\frac{%
\Delta \theta _{a}\left( E\right) }{\sqrt{U-E}}dE,  \label{inversasdeR}
\end{equation}
where the explicit dependency of the apsidal angle on the energy was stressed.  

If all bounded orbits are closed then the apsidal angle $\Delta
\theta _{a}\left( E\right) $, for these orbits, cannot change when the energy changes in a continual manner otherwise the continual changes would inevitably lead to open orbits. Taking this fact into account let us determine the central potentials that produce the same apsidal angle for all radially bounded orbits. After integrating equation (\ref{inversasdeR}) we obtain
\begin{equation}
\frac{1}{r_{1}}-\frac{1}{r_{2}}=\frac{2\sqrt{2m}\Delta \theta _{a}}{\ell\pi }%
\sqrt{U-U_{0}}  . \label{Resultado_tiko}
\end{equation}
Equation (\ref{Resultado_tiko}) was derived in Ref. \cite{tikochinsky} where a perturbative technique applied on a circular orbit leads to Bertrand's result.  The functions $r_{1}\left( U\right) $ and $r_{2}\left( U\right) $ being the inverse function of the function $U\left( r\right) $ are not independent of each other, and combined as they are in equation (\ref
{Resultado_tiko}), do not allow an efficient manipulation and hide the unique inverse we are looking for. At this point we perform an analytical continuation of the function $U\left(
r\right) $ such that we can consider its inverse function $r=r\left( U\right) $. Therefore we write 
\begin{equation}
\frac{1}{r}=\frac{\sqrt{2\mu}\Delta \theta _{a}}{\pi \ell }\sqrt{%
U-U_{0}}+\Phi \left( U,U_{0}\right) ,  \label{equ1/r}
\end{equation}
where $\Phi \left( U,U_{0}\right) $ is an analytical function of the complex variable  $U$ in an open neighborhood of $U_{0}$ satisfying the condition $\Phi \left( U_0,U_{0}\right)=1/r_0 $, and whose analytical continuation cannot have poles but can have other ramification points. Notice that it s not necessary to make use of the symbol $\pm$ before the second term of equation \ref{equ1/r}) because the square root has two branches. The positive sign corresponds to $r<r_{0}$ and the negative one to $r>r_{0}$. Taking equation (\ref{potencial_eft}) into equation (\ref
{equ1/r}) we obtain
\begin{equation}
\frac{1}{r}=\frac{\sqrt{2m}\Delta \theta _{a}}{\ell\pi }\frac{1%
}{r}\sqrt{r^{2}V\left( r\right) +\frac{l^{2}}{2\mu}-U_{0}r^{2}}+\Phi \left(
U,U_{0}\right)  . \label{identidade}
\end{equation}
The left-hand side of the identity $\left( \ref{identidade}\right) $ represents a meromorphic function with a single pole at $r=0$ and the right hand side of this same identity contains several terms but only one can spoil the analyticity of the complete function at some point not equal to $r=0$, namely the term that depends on the square root that generates a branch point at $r=r_{0}$. To avoid this it is mandatory to undo the branching effect inherent to the square root. This is possible only if the radicand is the square of an analytical function with a zero at $r=r_{0}$. In this way we identify two possibilities for the potential $V\left( r\right)$, to wit
\begin{eqnarray}
V\left( r\right) &=&-\frac{\kappa}{r} , \text{ \qquad \quad newtonian potential}, \\
V\left( r\right) &=&\frac{1}{2}\,\kappa\,r^{2} ,\text{ \qquad \text{isotropic harmonic oscillator potential}};
\end{eqnarray}
for which the apsidal angle is independent of the energy. We can calculate the corresponding constant apsidal angles for those two potentials as follows. 

For the newtonian potential the effective potential, equation (\ref{potencial_eft}), is given by
\begin{equation}
U=-\frac{\kappa}{r}+\frac{\ell^{2}}{2\mu r^{2}}.  \label{potefetivok}
\end{equation}
Solving equation (\ref{potefetivok}) with respect to  $1/r$ we obtain 
\begin{equation}
\frac{1}{r}=\frac{\mu \kappa}{\ell^{2}}+\frac{\sqrt{2\mu}}{l}\sqrt{U+\frac{\mu \kappa^{2}}{2l^{2}}%
} .  \label{sobrerk}
\end{equation}
Making use of equation (\ref{minimodeU}) with the effective potential given by   equation (\ref{potefetivok}) we find $r_{0}=\ell^{2}/(\mu \kappa) $ and the corresponding minimum energy  $U_{0}=-\mu \kappa^{2}/(2\ell^{2}) $. Now we can recast equation (\ref{sobrerk}) into the form 
\begin{equation}
\frac{1}{r}=\frac{1}{r_{0}}+\frac{\sqrt{2\mu}}{l}\sqrt{U-U_{0}} .
\label{sobrerku}
\end{equation}
Comparing equation (\ref{equ1/r}) with equation (\ref{sobrerku}) we can finally determine the apsidal angle for the newtonian potential which reads
\begin{equation}
\Delta \theta _{a}=\pi .
\end{equation}

The procedure employed with the newtonian potential can be also applied with a little bit more of effort to the case of the isotropic harmonic oscillator. The effective potential is now given by
\begin{equation}
U=\frac{1}{2}kr^{2}+\frac{\ell^{2}}{2\mu r^{2}} .
\end{equation}
This equation is a quartic equation in 
$1/r$, biquadratic more precisely,  and its solution is given by
\begin{equation}
\frac{1}{r}=\frac{1}{r_{0}}\sqrt{\frac{U}{U_{0}}+\sqrt{\left( \frac{U}{U_{0}}
\right)^{2}-1}} .  \label{sobrerou}
\end{equation}
Factoring out the right hand side of the equation (\ref{sobrerou}) we have
%
%$\allowbreak $%
\begin{equation}
\frac{1}{r}=\frac{\sqrt{2\mu}}{2l}\sqrt{U-U_{0}}+\frac{\sqrt{2\mu}}{2\ell}\sqrt{%
U+U_{0}},  \label{sobrerouf}
\end{equation}
where now we have made use of the relations $r_{0}^{2}=\ell/{\sqrt{\mu \kappa}}$ and $U_{0}=\ell\sqrt{\kappa/\mu}$. Comparing equations (\ref{equ1/r}) and (\ref{sobrerouf}) we obtain
\begin{equation}
\frac{\sqrt{2\mu}}{2\ell}=\frac{\sqrt{2\mu}\Delta \theta _{a}}{l\pi },\qquad
\therefore \qquad \Delta \theta _{a}=\frac{\pi }{2} .
\end{equation}
We can see that both potentials for which the apsidal angle is constant the orbits are closed. For the newtonian case the radius oscillates only once in a  complete cycle and for the oscillator case the radius oscillates  twice. 
\section{Final Remarks}
In this brief paper we derived Bertrand's theorem in a non-perturbative way. We have shown that simple analytical function techniques applied to the problem of finding the only central fields that allow an entire class of bounded, closed orbits with a minimum number of restrictions leads in a concise, straightforward way directly to the two allowed fields. We believe that the derivation discussed here is a valid alternative to a non-perturbative proof of Bertrand's theorem and can be presented at the undergraduate and graduate level or assigned as a problem for classroom discussion. 


\begin{thebibliography}{99}
%
\bibitem{bertrand} Bertrand J 1873 \textit{C.R. Acad. Sci. Paris } \textbf{77} 849

\bibitem{Newton} Newton I 1687 \textit{Philosophiae Naturalis Principia Mathematica} (London: Royal Society). English translation by A Motte revised by F Cajori 1962 (University of California Press, Berkeley CA)

\bibitem{tikochinsky} Tikochinsky Y 1988 \textit{Am. J. Phys.} \textbf{56} 1073 

\bibitem{Brown} Brown L S 1978 \textit{Am. J. Phys.} \textbf{46} 930

\bibitem{Zarmi} Zarmi Y 2002 \textit{Am. J. Phys.} \textbf{70} 446 

%\bibitem{Arnold} Arnol'd V I 1974 \textit{Mathematical Methods of Classical %Mechanics} (Springer: Berlim)

\bibitem{Arnold} Arnol'd V I 1976 \textit{Les M\'ethodes Math\'ematiques de la M\'ecanique Classique} (Mir: Moscou)

%\bibitem{Pesic98} Pesic P, 1998 \textit{Eur. J. Phys.} \textbf{19} 151

\bibitem{goldstein} Goldstein H, Poole C and Safko J 2002 \textit{ Classical Mechanics}  3rd edn (Reading: Addison-Wesley)

\bibitem{landau} Landau L and Lifchitz E 1969 \textit{ M\`{e}canique} 3$^{e}$
\`{e}dition revue (Mir: Moscou)

%\bibitem{landau} Landau L and Lifchitz E 19XX \textit{ Mechanics  (Pergamon: %London)


\bibitem{oldham} Oldham K B and Spanier J 1974 \textit{The Fractional Calculus} (London: Academic Press)

\end{thebibliography}
\end{document}